\documentclass[apj]{emulateapj}
\usepackage{apjfonts}

\newcommand*{\Msun}{\ensuremath{\mathrm{M_\odot}}}%
\newcommand*{\Mstar}{\ensuremath{M_\ast}}
\newcommand*{\pmt}{\ensuremath{\phi(\Mstar,t)}}
\newcommand*{\Mdot}{\ensuremath{\dot{M}_\ast}}
\newcommand*{\Mdotmt}{\ensuremath{\Mdot(\Mstar,t)}}

\newcommand*{\dd}{\mathrm{d}}
\newcommand*{\dpt}{\partial}

\newcommand*{\sigsfr}{\Sigma_{\mathrm{SFR}}}

\begin{document}


\title{The contribution of star formation and merging to stellar mass
  buildup in galaxies.}

\shorttitle{Stellar mass buildup in galaxies}
\shortauthors{Drory \& Alvarez}


\author{Niv Drory} \affil{Max-Planck Institut f\"ur extraterrestrische
  Physik (MPE), Giessenbachstrasse, Garching, Germany \\
   {\tt drory@mpe.mpg.de}}
\and
\author{Marcelo Alvarez} \affil{Kavli Institute for Particle Astrophysics
  and Cosmology, Stanford University, Stanford, CA 94309\\
  {\tt malvarez@slac.stanford.edu}}

\slugcomment{ApJ, in press}


\begin{abstract}
  We present a formalism to infer the presence of merging activity by
  comparing the observed time derivative of the galaxy stellar mass
  function (MF) to the change of the MF expected from the star
  formation rate (SFR) in galaxies as a function of galaxy mass and
  time. We present the SFR in the Fors Deep Field as a function of
  stellar mass and time spanning $9 < \log\Mstar < 12$ and $0<z<5$. We
  show that at $z \gtrsim 3$ the average SFR, $\Mdot$, is a power law
  of stellar mass ($\Mdot \propto \Mstar^{0.6}$). The average SFR in
  the most massive objects at this redshift are 100 -- 500~$\Msun
  \mathrm{yr}^{-1}$. At $z \sim 3$, the SFR starts to drop at the high
  mass end. As redshift decreases further, the SFR drop at
  progressively lower masses (downsizing), dropping most rapidly for
  high mass ($\log\Mstar \gtrsim 11$) galaxies. The mass at which the
  SFR starts to deviate from the power-law form (break mass)
  progresses smoothly from $\log \Mstar^1 \gtrsim 13$ at $z\sim 5$ to
  $\log \Mstar^1 \sim 10.9$ at $z\sim 0.5$. The break mass is evolving
  with redshift according to $\Mstar^1(z) = 2.7\times
  10^{10}\,(1+z)^{2.1}$.  We directly observe a relationship between
  star formation history (SFH) and galaxy mass.  More massive galaxies
  have steeper and earlier onsets of star formation, their peak SFR
  occurs earlier and is higher, and the following exponential decay
  has a shorter $e$-folding time. The SFR observed in high mass
  galaxies at $z \sim 4$ is sufficient to explain their rapid increase
  in number density. Within large uncertainties, we find that at most
  0.8 effective major mergers per Gyr are consistent with the data at
  high $z$, yet enough to transform most high mass objects into
  ellipticals contemporaneously with their major star formation
  episode. In contrast, at $z < 1.5$ and at $\Mstar \gtrsim 11$,
  mergers contribute 0.1 -- 0.2~Gyr$^{-1}$ to the relative increase in
  number density.  This corresponds to $\sim 1$ major merger per
  massive object at $1.5 > z > 0$.  At $10 < \log\Mstar < 11$, we find
  that these galaxies are being preferably destroyed in mergers at
  early times, while at later times the change in their numbers turns
  positive. This is an indication of the top-down buildup of the red
  sequence suggested by other recent observations.
\end{abstract}



\section{Introduction}\label{sec:introduction}

Galaxy formation and evolution in a cold dark matter (CDM) dominated
universe can be described in the most general terms as being driven by
the gravitationally induced hierarchical growth of structure in the
dark matter component coupled to the baryonic matter within the
merging and accreting dark matter halos. While the dark matter is
collisionless, baryonic matter dissipates energy by radiative cooling
and thereby is able to form stars.  Feedback effects are thought to
self-regulate the latter process.

Directly observing the fundamentally important hierarchical growth of
dark matter halos is not possible at the present time. Information
about the growth of the halos has to be deduced from observations of
galaxies or clusters. Large gravitational lensing surveys detecting
cosmic shear will allow the study of dark matter structure formation
directly in the future. In the meantime, it has proven difficult to
deduce halo growth from observations of galaxy growth.

Galaxies themselves grow in stellar mass firstly by star formation,
and secondly by accreting smaller galaxies and by merging with
similarly sized galaxies. Both of these processes need to be mapped
out if we wish to fully understand the assembly history of galaxies
and their host dark matter halos.

Progress has been made in recent years in measuring the build-up of
stellar mass density from redshift $z \sim 6$ to the present epoch
\citep{BE00,MUNICS3,Cohen02,DPFB03,Fontanaetal03,Rudnicketal03,%
  GDDS2,CBSI05,Rudnicketal06,EBELSSC07,Grazianetal07,SBEEL07}.  Also,
the stellar mass function has been mapped locally
\citep{2dF01,BMKW03,PerezGonzalez03a} and in some detail to $z \sim
1.5$
\citep{MUNICS6,K20-04,Bundyetal05,Borchetal06,Panellaetal06,Bundyetal06,Pozzettietal07}
with generally good agreement between different datasets. To some
lesser detail and accuracy deep surveys have provided data spanning $0
< z \lesssim 5$
\citep{FDFGOODS05,CBP05,Fontanaetal06,Yanetal06,Grazianetal07} and
even some estimates at $z \sim 7$ \citep{Bouwensetal06}. These mass
functions tend to agree fairly well at $z \lesssim 2$, while the
integrated mass densities may differ by a factor of $\sim 2$ at higher
$z$ (see, e.g., comparisons in \citealp{FDFGOODS05} and
\citealp{Fontanaetal06}). In summary, data are now available that
describe the mass in stars from an early epoch where only $\sim 1\%$
of the current mass density was in place until the present time.

Deep optical and near-infrared surveys, as well as mid-infrared data
from the Spitzer Space Telescope and sub-mm surveys have greatly
expanded our understanding of the cosmic star formation (SF)
history. The integrated star formation rate \citep{Madauetal96} is now
quite well constrained to $z \sim 4$ (with some estimates at even
higher $z$) from a multitude of studies employing different methods at
wavelengths from the UV to the far-IR
\citep{CFRS96,MPD98,Floresetal99,PFH99,Kauffmannetal03a,HPJD04,%
  Gabasch04a,CBSI05,Belletal05,WCB06,Thompsonetal06,HB06,NOFC06,%
  Yanetal06,Bouwensetal06}.

In the recent literature, more attention is being devoted to the star
formation rate (SFR) as a function of stellar mass, usually in the
form of the specific star formation rate (SSFR) as a function of
stellar mass
\citep{CSHC96,BE00,Fontanaetal03,PerezGonzalez03b,Brinchmannetal04,%
  BDHF04,Juneauetal05,Salimetal05,Feulneretal05,Feulneretal05b,%
  PerezGonzalez05,FHB06,Belletal07,Zhengetal07}.

Progress on the merger rate and its evolution with redshift has been
somewhat slower. Observational results obtained from galaxy pair
statistics vary considerably, finding merger rates $\propto (1+z)^m$
with $0 < m \lesssim 4$ in the redshift range $0<z \lesssim 1$
\citep{ZK89,BKWF94,YE95,NIRGC97,LeFevreetal00,Carlbergetal00b,%
  Pattonetal01,Linetal04,Bundyetal04,DePetal05,Conselice06,%
  Belletal06b,Kartaltepeetal07}.  Early studies tend to give somewhat
higher numbers than the more recent work. The diversity of these
observational merger rates is likely to be due to differing
observational techniques, differing criteria for defining a merging
pair, and survey selection effects. To this date, no results have been
published for significantly higher redshifts.  Note that Cold Dark
Matter (CDM) N-body simulations predict halo merger rates $\propto
(1+z)^m$ with $2.5\lesssim m \lesssim 3.5$
\citep{Governatoetal99,Gottloberetal01,Gottloberetal02}.

Little is known about the dependence of the merger rate on galaxy
luminosity or mass. Recently, \cite{XSH04} find evidence from the
$K_\mathrm{s}$-band luminosity function of galaxies in major-merger
pairs that the local merger rate decreases with galaxy mass.

In this paper, we present a method to infer the presence of merging
activity and as such the underlying hierarchical growth of structure
from a comparison of observations of galaxy mass functions with
observations of star formation rates in the galaxies. Essentially, we
compare a measurement of the time derivative of the mass function (the
number density of galaxies per unit stellar mass; $\pmt$) with an
expectation derived from time-integrating the (average) star formation
rate of galaxies as a function of mass, $\Mdotmt$. The difference
between the two can then be attributed to mass function evolution due
to ``external causes'', which we identify with accretion and merging
(after taking the survey selection effects into account). We present
our results with caution, given the limitations of the current
dataset.

This paper is organized as follows: In \S~\ref{sec:desc-mass} we
establish a formalism that compares the change in the stellar mass
function expected from the observed star formation rate as a function
of mass and time to the observed time derivative of the mass
function. The aim is to deduce a merger rate (or, more precisely, the
change in number density due to merging) as a function of galaxy mass
and time from the difference of the two. We will call this quantity
the {\em assembly rate}. \S~\ref{sec:mass-function} and
\S~\ref{sec:sfr} present the mass function data and the star formation
rate data, respectively. Also, suitable parametrizations of the data
and its evolution with time are presented and
discussed. \S~\ref{sec:dphidt} discusses the total time derivative of
the mass function. \S~\ref{sec:dphisfr} discusses the change of the
mass function that is expected due to the star formation rate
only. \S~\ref{sec:mergerrate} finally combines the former two results
to discuss the change in number density due to merging we infer from
our data. All these are summarized and discussed in context in
\S~\ref{sec:summary}.

Throughout this work we assume $\Omega_M = 0.3$, $\Omega_{\Lambda} =
0.7, H_0 = 70\ \mathrm{km\ s^{-1}\ Mpc^{-1}}$.


\section{Describing the Evolution of Galaxy Mass}\label{sec:desc-mass}

A general description of the evolution of galaxy stellar mass with
time can be given in terms of (1) internal evolution due to star
formation in the galaxy, and (2) externally-driven evolution due to
the infall of other smaller galaxies (accretion) or the merging with
another galaxy of comparable size. The former can be measured by
finding the average star formation rate as a function of stellar mass
and time, $\Mdotmt$. The latter cannot be measured reliably at this
point. However, if one knew the time derivative of the galaxy stellar
mass function (GSMF), $\dpt \phi(\Mstar,t) / \dpt t$, then it could in
principle be derived by comparing the internal evolution due to star
formation to the total observed evolution, and thereby inferred from
the difference between the two.

First, we note that the evolution of the GSMF due to star formation in
the galaxies by definition preserves the number of galaxies. Hence the
change of the GSMF due to star formation is described by a continuity
equation in $\pmt$:

\begin{equation}
  \label{eq:cont}
  \frac{\dpt \pmt}{\dpt t} \Big\vert_{\mathrm{SF}} \; = \;
  - \frac{\dpt}{\dpt \Mstar} \left[ \pmt \, \Mdotmt \right]\, .
\end{equation}

Here, somewhat unconventionally, we have taken $\vert_{\mathrm{SF}}$
to mean ``due to star formation''.

This equation becomes intuitively clear if one considers that the
continuity equation above means that the change in numbers in a given
stellar mass bin $[M,M+dM)$ is given by the difference across this bin
of the number of objects entering the bin from below and the number of
galaxies leaving the bin to higher masses, both owing to their star
formation rate. Hence the derivative with respect to mass of the
number density times the star formation rate (which gives the
difference of mass flux across the bin) on the right hand side.

The change with time of the GSMF, $\dpt \phi(\Mstar,t) / \dpt t$, and
the average SFR as a function of galaxy mass and time, $\Mdotmt$, are
in principle directly observable. We can therefore infer the change in
the GSMF due to all processes other than star formation (which we will
identify with merging) from these two functions. We will call the
resulting quantity the {\em assembly rate}:

\begin{equation}
  \label{eq:merge}
  \frac{\dpt \pmt}{\dpt t} \Big\vert_{\mathrm{Merge}} \; = \;
  \frac{\dpt \pmt}{\dpt t} \, - \,
  \frac{\dpt \pmt}{\dpt t} \Big\vert_{\mathrm{SF}}\, .
\end{equation}

Hence,

\begin{equation}
  \label{eq:merge2}
  \frac{1}{\phi}\frac{\dpt \phi}{\dpt t} \Big\vert_{\mathrm{Merge}} \; = \;
  \frac{1}{\phi}\frac{\dpt \phi}{\dpt t} \, + \,
  \frac{1}{\phi}\frac{\dpt}{\dpt \Mstar} \left[ \phi \, \Mdot \right]\,
\end{equation}

\begin{figure*}[t]
  \centering
  \includegraphics[width=0.9\textwidth]{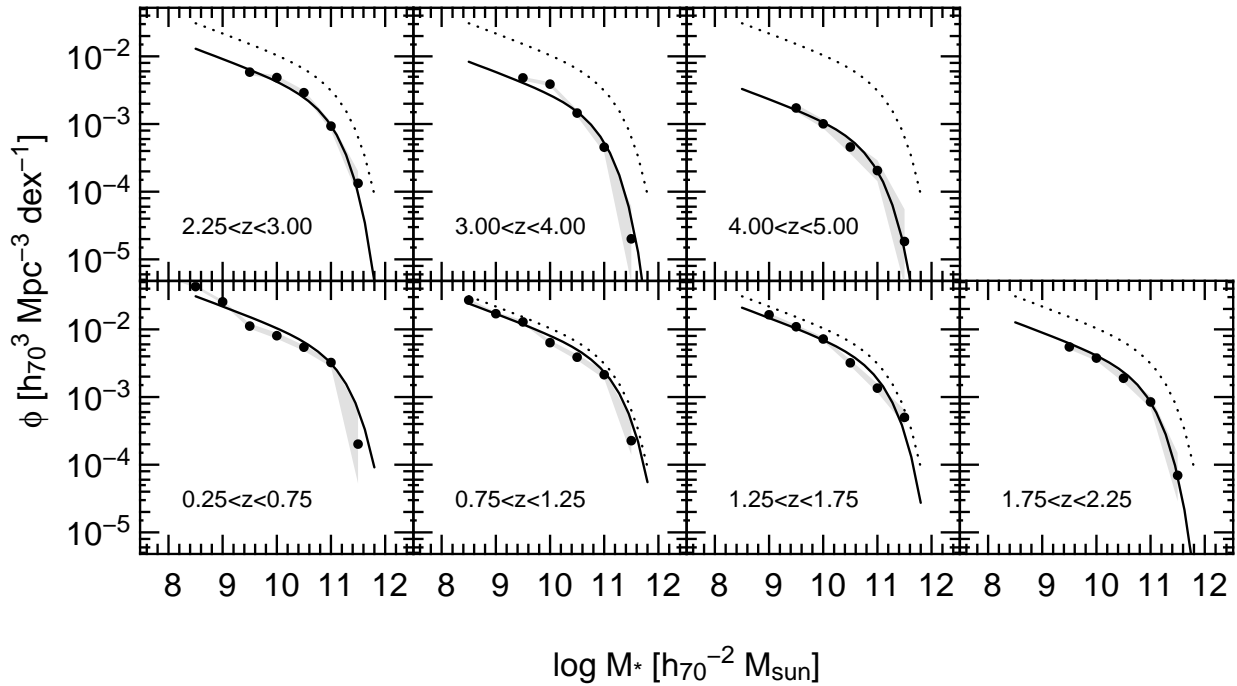}
  \caption{The stellar mass function (MF) as a function of
    redshift. Data are from the Fors Deep Field
    \protect\citep{FDFGOODS05}. Schechter-fits to the data are over plotted as
    solid lines. Confidence regions are indicated by shading (random
    uncertainties only). The low-$z$ MF is repeated in the higher-$z$
    panels for comparison.\label{fig:mf}}
\end{figure*}

where we have left out the arguments for brevity and divided by $\phi$
to obtain an equation for the relative change.

Note that (tidal) stripping of stars from a galaxy (e.g.\ in clusters
or in interactions) contributes to eq.~\ref{eq:merge2} as
well. Assuming that this loss of stellar mass is small compared to the
overall growth rate of a galaxy, it can be neglected.

The {\em assembly rate} is given in terms of the change in number
density per unit time rather than in number of discrete merging events
per object per unit time (and per unit merger mass ratio). Therefore,
we can only speak of an effective merger rate, equivalent in its
effect on the number density to an equal number of same-mass mergers
per unit time interval. An alternative is to think about this quantity
as simply representing the average effective mass accretion rate. We
cannot discern the mass spectrum of accretion and merging events that
actually cause the observed change in the mass function. It is also
important to point out that a value of zero does not imply that no
accretion and merging are happening, but only that the mass flux
entering a mass bin through merging and the mass flux leaving that
same mass bin through merging are equal, and hence the number density
at this mass does not change. Values below zero imply that the number
density decreases due to merging activity, and hence objects are
preferably being accreted onto more massive objects and thus
destroyed. To emphasize the difference between what we measure and the
merger rate expressed as the number of merger events per unit time and
object, we will avoid using the term merger rate and prefer the name
{\em assembly rate}.

In the following sections we shall compile the necessary terms from
GSMF and star formation data available in the literature. We will
concentrate on the methodology and defer discussing selection effects
and other deficiencies in the data to a later section where we proceed
to plot eq.~(\ref{eq:merge2}) and present our overall results.


\section{The Galaxy Stellar Mass Function}
\label{sec:mass-function}

The time derivative of the galaxy stellar mass function can be
obtained in a fairly straight-forward way from observational studies
of the GSMF. Here, we use the stellar mass function data spanning $0 <
z < 5$ by \citet{FDFGOODS05} from the Fors Deep Field (FDF).

The FDF photometric catalog is published in \citet{FDF1} and we use
the I-band selected subsample covering the deepest central region of
the field as described in \citet{Gabasch04}.  This catalog lists 5557
galaxies down to $I \sim 26.8$. The latter paper also shows that this
catalog misses at most 10\% of the objects found in ultra-deep K-band
observations found by \citet{FIRES03}. Distances are estimated by
photometric redshifts calibrated against spectroscopy up to $z \sim 5$
\citep{FDF2} providing an accuracy of \mbox{$\Delta z / (z_{spec}+1)
  \le 0.03$} with only $\sim 1$\% outliers \citep{Gabasch04}. The
spectroscopy samples objects statistically down to $I = 25$.

The stellar masses are computed by fitting models of composite stellar
populations of varying star formation history, age, metallicity, burst
fractions, and dust content to UBgRIZJK multicolor photometry.  We
assume a Salpeter stellar initial mass function. At faint magnitudes,
the uncertainties in the photometric distances are dominated by the
large uncertainties in the colors. These are taken into account in the
fitting procedure and hence reflected in the uncertainties in the
individual masses entering the mass function. We refer the reader to
the original publications for full details.

To lessen the impact of noise in the data on the time derivatives we
are interested in, we choose to parameterize the evolution of the mass
function, which we assume to be of Schechter form, by the following
relations:
\begin{equation}
  \label{eq:schechter}
  \nonumber
  \phi(M)\,\dd M \; = \;
  \phi^* \, \left(\frac{M}{M^*}\right)^\alpha \,
  \exp\left(-\frac{M}{M^*}\right)\frac{\dd M}{M^*}\, ,
\end{equation}
and
\begin{eqnarray}
  \label{eq:mf-parametrization}
  \phi^*(z)   & \; = \; & p_1\, (1+z)^{p_2}\, ,   \\
  \log M^*(z) & \; = \; & q_1 + q_2\,\ln(1+z)\, , \\
  \alpha(z)   & \; = \; & \alpha_0 \, .
\end{eqnarray}

\begin{figure}[t]
  \includegraphics[]{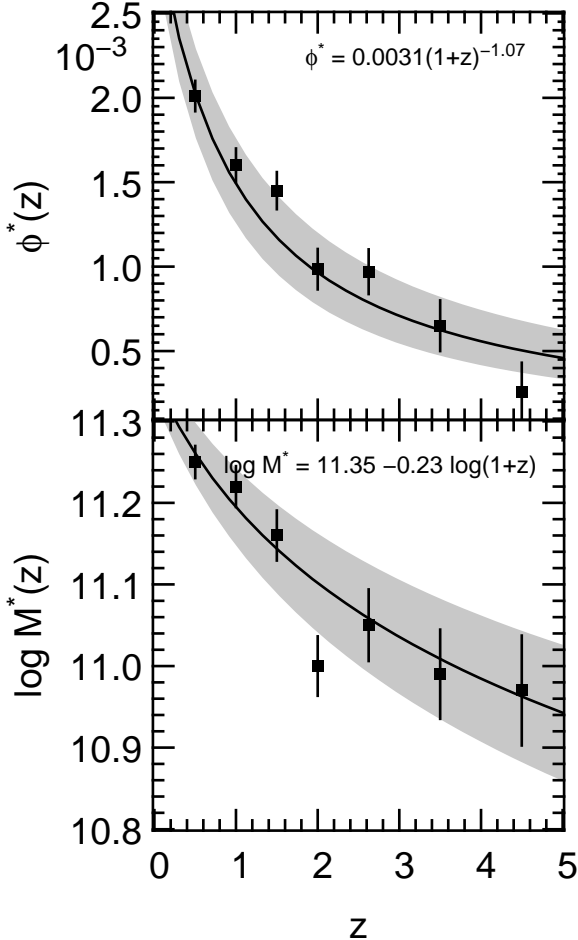}
  \caption{Evolution with redshift of the parameters $\phi^*$ and
    $M^*$ of the Schechter functions fit to the data shown in
    Fig.~\ref{fig:mf}. The shaded areas show the 1-$\sigma$ confidence
    region of the fit as determined from Monte Carlo simulations (see
    text). Note that the derivatives w.r.t.\ $z$ have much smaller
    uncertainties than the values of $\phi^*(z)$ and $M^*(z)$
    themselves.\label{fig:schechter}}
\end{figure}

First, we fit individual Schechter functions to the data from
\citet{FDFGOODS05} in each redshift bin (Fig.~\ref{fig:mf}). Then we
fit eqs.~(5)--(7) to the Schechter parameters as a function of
redshift, $\phi^*(z)$ and $\log M^*(z)$, to obtain a smooth
description of the redshift evolution of the stellar mass function
(Fig.~\ref{fig:schechter}). The error bars of each data point are used
to weight the data during fitting. We determine confidence intervals
for $\phi^*(z)$ and $\log M^*(z)$ by Monte Carlo simulations. These
confidence regions are indicated by shaded regions in
Fig.~\ref{fig:schechter}. A consequence of parameterizing $\phi^*(z)$
and $\log M^*(z)$ in this way is that in spite of the large
uncertainties of these quantities, the constraints on their
derivatives with respect to time are a lot tighter.

In the redshift range $0<z\lesssim 2$, where the mass function data
are deep enough to constrain the faint-end slope of the mass function,
$\alpha$, it turns out to be consistent with being constant at $\alpha
= -1.3$. At higher redshift, the data do not allow to constrain the
faint-end slope any more. This is consistent with the results by
\citealp{Fontanaetal06}.

We therefore fix alpha at its low-$z$ value
for all redshifts. The consequences of this will be discussed below.

The mass function and the Schechter-fits to the data are shown in
Fig.~\ref{fig:mf}, and the best-fitting Schechter parameters are
listed in Table~\ref{tab:mffit}.  The fit to the evolution of the mass
function with redshift as parameterized by eqs.~(5)--(7) is shown in
Fig.~\ref{fig:schechter}. The best fitting parametrization is
\begin{eqnarray*}
  \label{eq:mf-params}
  \phi^*(z)   & \; = \; & (0.0031\pm 0.0002)\, (1+z)^{-1.07\pm 0.13}\, , \\
  \log M^*(z) & \; = \; & (11.35\pm 0.04) - (0.22\pm 0.03)\,\ln(1+z) \, , \\
  \alpha(z)   & \; = \; & -1.3 \, ,
\end{eqnarray*}
and we will use this form in the following.

\begin{deluxetable}{llll}
\tablecolumns{4}
\tablecaption{\label{tab:mffit}Schechter fit parameters to the mass function}
\tablehead{
\colhead{$z$} & \colhead{$\phi^*$} & \colhead{$\log M^*$} & \colhead{$\alpha$}\\
\colhead{}    & \colhead{$h_{70}^3\,\mathrm{Mpc}^{-3}\,\mathrm{dex}^{-1}$}
              & \colhead{$h_{70}^{-2}\,\mathrm{M_\sun}$} & \colhead{}}
\startdata
  0.50 & $(2.01\pm 0.10)\times 10^{-3}$ & $11.25\pm 0.02$ & -1.3\\
  1.00 & $(1.60\pm 0.12)\times 10^{-3}$ & $11.22\pm 0.03$ & -1.3\\
  1.50 & $(1.45\pm 0.14)\times 10^{-3}$ & $11.16\pm 0.03$ & -1.3\\
  2.00 & $(9.85\pm 1.12)\times 10^{-4}$ & $11.00\pm 0.04$ & -1.3\\
  2.75 & $(9.70\pm 1.32)\times 10^{-4}$ & $11.05\pm 0.04$ & -1.3\\
  3.50 & $(6.50\pm 1.50)\times 10^{-4}$ & $10.99\pm 0.05$ & -1.3\\
  4.50 & $(2.61\pm 1.71)\times 10^{-4}$ & $10.97\pm 0.07$ & -1.3\\
\enddata
\end{deluxetable}

Using this best-fitting parametrization of the evolution of the mass
function, we can extrapolate the values of the parameters to $z=0$. We
obtain $\phi^*(0) =
0.0031~h_{70}^3\,\mathrm{Mpc}^{-3}\,\mathrm{dex}^{-1}$, which is
consistent with the values obtained by \citet{2dF01} (0.003) and
\citet{BMKW03} (0.0034) when these are rescaled to the same initial
mass function and cosmology. The value we obtain for the
characteristic mass, $\log M^*(0) = 11.35$, is higher than the values
obtained in the mentioned literature (11.16 and 11.01, respectively);
however, our value of the faint-end slope, $\alpha$, is different
($-1.3$ in this work as opposed to $-1.1$ in \citealp{2dF01} and
$-1.18$ in \citealp{BMKW03}). If one takes the coupling between $M^*$
and $\alpha$ into account, a smaller value of $\alpha$ is degenerate
with a higher value of $M^*$, and therefore our values agree very well
with the literature at zero redshift, once ``corrected'' to the same
$\alpha$ along the lines of the degeneracy.

In any case, the focus of this paper is to establish a method of
disentangling star formation and merging contributions to galaxy
stellar mass assembly. We shall therefore not be concerned too much
about peculiarities of the mass function, instead we refer the
interested reader to the relevant literature \citep[e.g.][and
references therein]{BMKW03,FDFGOODS05,CBP05,Fontanaetal06}.


\section{The Average Star Formation Rate}
\label{sec:sfr}

\begin{figure*}
  \centering
  \includegraphics[width=0.933\textwidth]{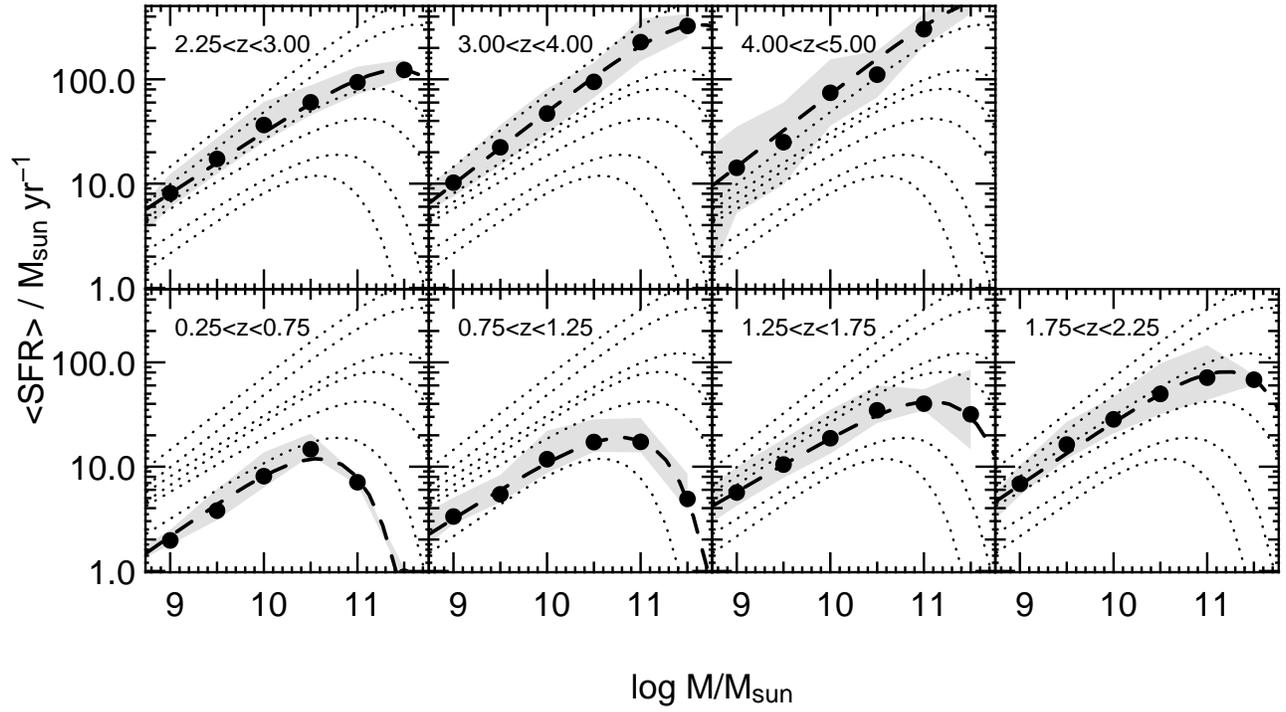}\vspace*{1cm}
  \includegraphics[width=0.7\textwidth]{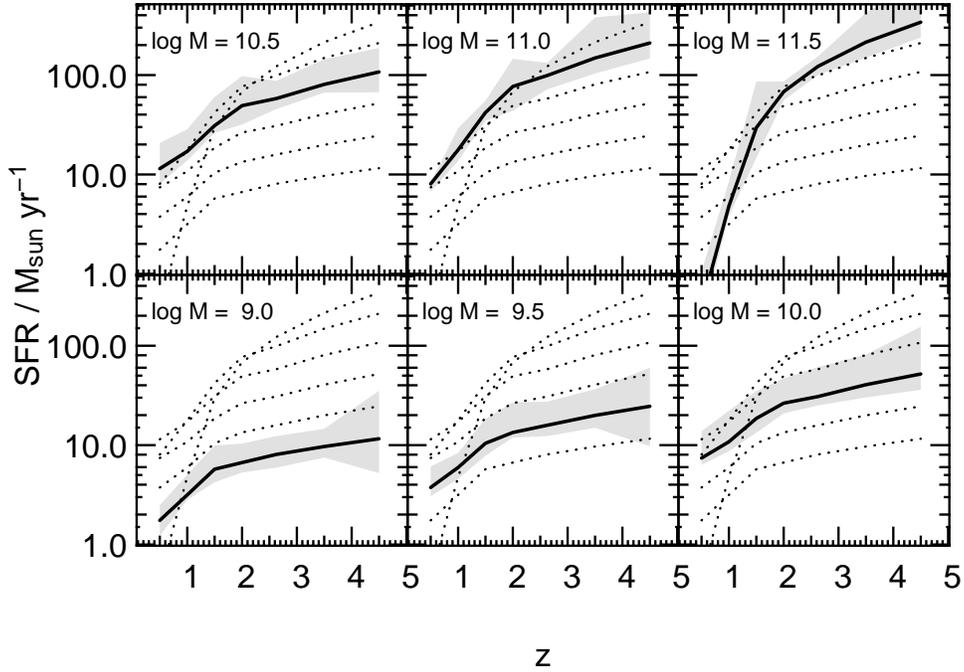}
  \caption{The {\em upper plot} shows the average star formation rate
    as a function of stellar mass in seven redshift bins ranging $0.5
    < z < 4.5$. Each panel shows the data of one redshift bin (filled
    circles).  The confidence regions are indicated by shading.
    Fitting functions to the data (power laws with exponential cutoff
    at high mass; see text) are plotted as dashed lines. Each panel
    also shows the fitting functions in all other redshift bins as
    dotted lines for comparison. The {\em lower plot} shows the same
    data plotted against redshift. Each panel shows the star formation
    rate for a different stellar mass as a thick solid line. The
    dotted lines indicate the data for the other masses for comparison.
    Confidence regions are shaded.
    \label{fig:sfr}}
\end{figure*}

To find the (average) star formation rate at each stellar mass and
time, we take advantage of the data published by \cite{Feulneretal05}
in the Fors Deep Field.

We use dust-corrected UV-continuum emission around 2800\AA\ to
estimate star formation rates. The reason for doing so is
twofold. First, there are no other homogeneous datasets that provide
more robust star formation rates over the whole redshift range of
interest. Second, the UV continuum luminosity can be measured for
every galaxy that has a stellar mass assigned to it. Therefore we have
star formation rates for the exact same sample that is used to measure
the stellar mass function. This at least provides some assurance of
internal consistency.

\begin{deluxetable}{llll}
\tablecolumns{4}
\tablecaption{\label{tab:sfrfit}Fit parameters to the star formation rate (Eq.~\ref{eq:sf-parametrization})}
\tablehead{
\colhead{$z$} & \colhead{$\Mdot^0$} & \colhead{$\log \Mstar^1$} & \colhead{$\beta$}\\
\colhead{}    & \colhead{$\Msun \, \mathrm{yr}^{-1}$}
              & \colhead{$\Msun$} & \colhead{}}
\startdata
  0.50 & $13.1 \pm 1.6$  & $10.77\pm 0.08$ & $0.69\pm 0.11$ \\
  1.00 & $20.1 \pm 1.0$  & $11.04\pm 0.04$ & $0.56\pm 0.08$ \\
  1.50 & $43.6 \pm 2.8$  & $11.35\pm 0.09$ & $0.52\pm 0.07$ \\
  2.00 & $88.2 \pm 3.1$  & $11.41\pm 0.11$ & $0.61\pm 0.07$ \\
  2.75 & $132  \pm 3.9$  & $11.67\pm 0.13$ & $0.59\pm 0.05$ \\
  3.50 & $336  \pm 12.7$ & $11.78 (-0.25 +\infty)$\tablenotemark{1} & $0.63\pm 0.02$ \\
  4.50 & $576  \pm 26.7$ & $12.95 (-0.31 +\infty)$\tablenotemark{1} & $0.66\pm 0.01$
\enddata
\tablenotetext{1}{Values are unconstrained on the high
  side. They are consistent with a single power law with no break at
  all.}
\end{deluxetable}

The dust corrections are estimated from the stellar population model
fits to the multicolor photometry (see details and discussion in
\citealp{Feulneretal05}).

How reliable are the UV-based star formation rates?
\citet{Daddietal07} analyze multi-wavelength data (optical-NIR, {\em
  Spitzer} mid-IR, and radio) at $1.4 < z < 2.5$ in the GOODS region
(see also \citealp{Daddietal05b}). They conclude that the SFR can be
estimated consistently from the UV light using SED or color-based dust
corrections, with the UV-based estimates being within a factor of
$\sim 2$ of the $24~\mu \mathrm{m}$-based estimate for most galaxies (Sub-mm
galaxies are an exception, as for those, the UV underestimates the SFR
by at least an order of magnitude, but these are likely missed by the
I-band selected sample used here). At high IR luminosities the
difference is higher, but it is often unclear whether it is due to an
IR excess due to an AGN, which is likely because the UV estimate is in
agreement with the radio-based estimate of the SFR. These conclusions
are also supported by a similar analysis by \citet{Reddyetal05}. These
data only probe relatively massive galaxies, though.

To estimate errors (or rather the range of possible values) of the
average star formation rate as a function of mass, we proceed as
following: instead of applying a further upward correction to all our
star formation rates, we set the likely upper envelope of the average
star formation rate as the average value plus the width of the
distribution above the average (using the 68-percentile). The
reasoning behind this is that the brightest UV sources at each mass
are likely the least extinct ones, and therefore they mark the upper
limit. The likely lower confidence limit on the average star formation
rate is calculated by estimating the maximum amount by which the
average might be lowered due to low-SFR galaxies missing in the
sample. To do this we need to assume a distribution of star formation
rates in objects below our detection limit. We assume a star formation
rate function power-law slope of $-1.45$ following the only
measurement by \citet{Belletal07}. Note, though, that this measurement
is at $z \lesssim 1$. We normalize to our mean densities and assume
that the undetected population follows a power-law distribution of
star formation rate of that slope at any given constant stellar mass
and redshift. We then integrate the star formation rate function up to
the point at which the objects would reach the detection limit of the
FDF. The amount by which this missed star formation lowers the average
value at each mass and redshift then defines the lower confidence
limit. The region of confidence is shaded in Fig.~\ref{fig:sfr}.

To reduce the influence of noise in the data on the derivatives we are
interested in, we will parameterize the average star formation rate as
a function of mass by an analytic expression, just as we have done
with the mass function.  We choose a power law of the stellar mass
with an exponential cutoff at high mass:
\begin{equation}
  \label{eq:sf-parametrization}
  \Mdot(\Mstar) \; = \;
  \Mdot^0 \, \left(\frac{\Mstar}{\Mstar^1}\right)^\beta \,
  \exp\left(-\frac{\Mstar}{\Mstar^1}\right)\, .
\end{equation}

The data along with this fit are shown in Fig.~\ref{fig:sfr}.  The
best-fitting parameters are listed in Table~\ref{tab:sfrfit}.  Since
such data on the star formation rate as a function of stellar mass and
time have not been shown before in this form in the literature, we
will take the time to digress from our goal of determining the assembly
rate and discuss the star formation rate as it provides valuable
insights.

A few regularities in the data are apparent. At $z \gtrsim 3$ the star
formation rate monotonically rises with stellar mass to the highest
masses we probe ($\log \Mstar \sim 11.5$). It is consistent with a
single power law with exponent $\beta \sim 0.6$. The star formation
rates in the most massive objects at this high redshift are of the
order of 100 -- 500~$\Msun \mathrm{yr}^{-1}$, suggesting that these
are the progenitors of modern day massive ellipticals seen during
their major epoch of star formation. The average star formation rate
drops most rapidly with redshift for high mass ($\log\Mstar \gtrsim
11$) galaxies. Their period of active growth by star formation seems
to be over by $z \sim 1.5$, and the star formation rate as a function
of stellar mass turns over from increasing with mass to rapidly
decreasing with both mass and time in objects with $\log\Mstar \gtrsim
11$.

At redshift $z \sim 3$, the star formation rates begin to drop at the
high mass end while still being an increasing function of mass at
lower masses. As redshift decreases further, the star formation rates
drop at progressively lower masses. This behavior, usually termed
downsizing \citep{CSHC96}, has been observed in numerous recent
high-redshift studies
\citep[e.g.][]{BDHF04,VdWetal05,Bundyetal06,Noeskeetal07b} and is also
seen in studies of the fossil record of the stellar populations in the
local universe \citep[e.g.][]{HPJD04,TMBM05}. The mass at which the
star formation rate starts to deviate from the power law form
progresses smoothly from $\log \Mstar^1 \sim 13$ at $z\sim 5$ to $\log
\Mstar^1 \sim 10.5$ at $z\sim 0.5$. For comparison, in the local
universe \citet{Kauffmannetal03b} find a characteristic mass of $\log
\Mstar \sim 10.4$ at which the galaxy population changes from actively
star forming at lower stellar mass to typically quiescent at higher
masses. This transition mass shows a smooth evolution to higher masses
at earlier times up to very high redshift (at $ \gtrsim 3$ the data
are consistent with a single power law and no break mass at all).

\begin{figure}[t]
  \centering
  \includegraphics[width=8cm]{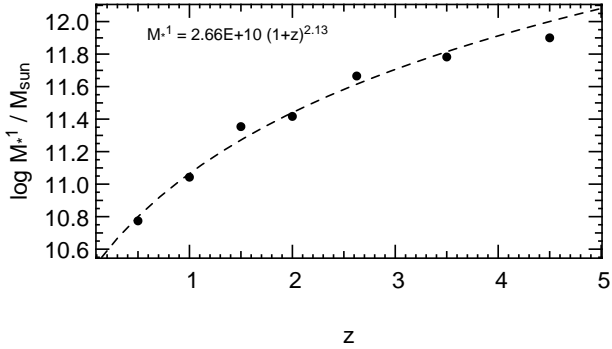}
  \caption{Evolution of the characteristic mass, $\Mstar^1$, above
    which the star formation rate as a function of mass begins to
    drop.\label{fig:mq}}
\end{figure}
How does this ``break mass'' -- the mass above which the average star
formation rate begins to decline -- evolve with redshift?
Fig.~\ref{fig:mq} shows the evolution of the parameter $\Mstar^1$ from
eq.~(\ref{eq:sf-parametrization}) against redshift, as well as a
power-law fit to it:
\begin{equation}
  \Mstar^1(z) = 2.7\times 10^{10} (1+z)^{2.1}\, .
\end{equation}

This can be compared directly to the result obtained from the SDSS
sample in the local universe by \cite{Kauffmannetal03b} (compare also
\citealp{Brinchmannetal04}). They find that above a stellar mass of
$3\times 10^{10}~\Msun$, the fraction of galaxies with old stellar
populations and low star formation rapidly increases. Our estimate of
the zero redshift break mass in the star formation rate is
$\Mstar^1(0) = 2.7\times 10^{10}~\Msun$. \citet{Bundyetal06} find
evidence for an evolving mass limit for star forming galaxies in the
DEEP2 redshift survey at $0.4 < z < 1.4$. Their best-fitting
parametrization of the ``quenching mass'', $M_Q(z) \propto
(1+z)^{3.5}$, is probably comparable to ours given the difference in
methodology and redshift range, although the exponent we find is
somewhat smaller.

In the lower panels of Fig.~\ref{fig:sfr}, we plot the star formation
rate against redshift for different fixed stellar masses. From this
plot it is apparent that while massive galaxies have the highest
(absolute) star formation rates, their star formation rates drop
earlier and quicker than for lower stellar mass objects. The star
formation rate in $\log\Mstar \sim 11$ objects drops by a factor $\sim
200$ between $z\sim 4.5$ and $z\sim 1$, while objects of $\log\Mstar
\sim 9.5$ decrease their average star formation rate only by a factor
$\sim 5$. Most of the latter decrease happens at $z \lesssim 1$. At
higher redshifts, low mass galaxies have almost constant (average)
star formation rates. The more rapid decline we find in the star
formation rate of massive galaxies is in contrast with the analysis by
\citet{Zhengetal07} who combine $24~\mu\mathrm{m}$\ data with UV data
in the COMBO-17 survey and conclude that the star formation rate
declines independently of galaxy mass at $z<1$ (using stacked
detections to account for the average star formation rate in sources
individually undetected at $24~\mu\mathrm{m}$). This difference could
be due to the (SED-based dust corrected) UV light underestimating the
true star formation rate in $\log \Mstar \gtrsim 11$ galaxies at $z<1$
more so than at lower masses. Another possibility is some uncertainty
in the infrared SEDs of these galaxies and the calibration of the
observed-frame $24~\mu\mathrm{m}$-based star formation rate at
intermediate redshifts.

Another fact to note is that the exponent of the star formation rate
as a function of stellar mass at masses below the break mass is
remarkably stable. In all redshift bins, the exponent is close to
$\beta \sim 0.6$, possibly steepening slightly to $\beta \sim 0.7$ at
$z \lesssim 0.5$. This also means -- since the exponent is always
smaller than one -- that the specific star formation (star formation
rate per unit stellar mass) rate is a monotonically decreasing
function of stellar mass at all masses and redshifts.

How does this compare to other studies of star formation rates? In the
local universe, \citet{Brinchmannetal04} find a very similar slope of
the star formation rate with stellar mass as well ($\log \Mdot \propto
0.6\, \log \Mstar$).

At intermediate redshift, in fact, \citet{Noeskeetal07a} argue for the
existence of a ``main sequence'' of star forming galaxies that obeys
the relation $\log \Mdot = (0.67\pm 0.08)\, \log \Mstar - (6.19\pm
0.78)$ at $0.2 < z < 0.7$ and $10 < \log \Mstar/\Msun < 11$ using
DEEP2 survey data combining $24~\mu \mathrm{m}$\ and optical emission
line fluxes. This is in very good agreement with our results. They
also see indications of slight flattening towards higher redshift,
although this cannot be said with confidence from their sample.

\citet{Papovichetal06} studied massive galaxies at $1 \lesssim z
\lesssim 3.5$ detected by the Spitzer Space Telescope at
$3-24~\mu\mathrm{m}$, by HST, and in ground based observations in the
GOODS field. They also find a relation between stellar mass and star
formation rate, and although they do not attempt to quantify this
relation, their data is consistent with the relation proposed
here. They also compare their high-$z$ results to similarly measured
star formation rates in $z \lesssim 1$ galaxies selected from COMBO-17
and find similar results to the evolution presented here.

Also, our star formation rates (and stellar masses) in the Lyman break
galaxy regime agree very well with detailed studies of this class of
objects. As in \citet{PDF01}, we find stellar masses of $9.5 \lesssim
\log\Mstar \lesssim 10.5$ and star formation rates (at $z \sim 3$)
between 10 and 100~$\Msun \mathrm{yr}^{-1}$.

We are therefore confident that our relationship between stellar mass
and star formation rate is robust and likely to be valid in spite of
its rather simplistic footing. It is understood that there is an
underlying broad distribution of star formation rates and much
intrinsic scatter around this relation, especially at higher redshift
where one is likely to encounter massive starbursts. Also, the
relation as shown here is only strictly valid for ``star forming
galaxies'', as quiescent objects (most notably at lower masses) are
most likely to be undetected (however, they will be missing from both
$\phi(\Mstar)$ and $\Mdot(\Mstar)$ in the present study).

\begin{figure}[t]
  \centering
  \includegraphics[width=8cm]{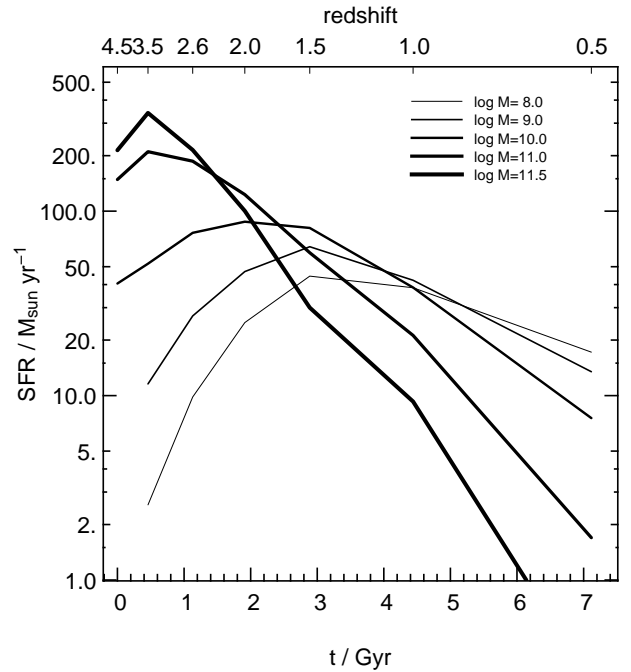}
  \caption{Star formation histories of galaxies with mass between
    $10^8$ and $10^{11.5}$~\Msun, calculated by integrating the star
    formation rate as a function of mass and time
    (Fig.~\ref{fig:sfr}). \label{fig:avgsfh}.}
\end{figure}

Finally, given the average star formation rate field
$\Mdot(\Mstar,t)$, we can ask what the star formation histories look
like on average as a function of mass, i.e.\ we can ask for tracks
along $\Mdot$ and $t$ that solve the equations
\begin{eqnarray}
  \nonumber
  \dd \Mstar/\dd t & \; = \; & \Mdotmt\\
  \nonumber
  \Mstar(t_0) & \; = \; & M_0\, .
\end{eqnarray}
Figure~\ref{fig:avgsfh} shows these solutions for masses $M_0$ at
$t_0$ of $8 \leq \log\Mstar \leq 11.5$. Note that $M_0$ and $t_0$ are
defined as the mass and time at $z=4.5$.  In interpreting these as
star formation histories of galaxies, we assume that the dominant
contribution to the growth of stellar mass is from star formation,
i.e.\ we neglect growth by accretion and merging, which we will see is
a reasonable approximation in the following sections.

\begin{figure*}[t]
  \centering
  \includegraphics[width=0.45\textwidth]{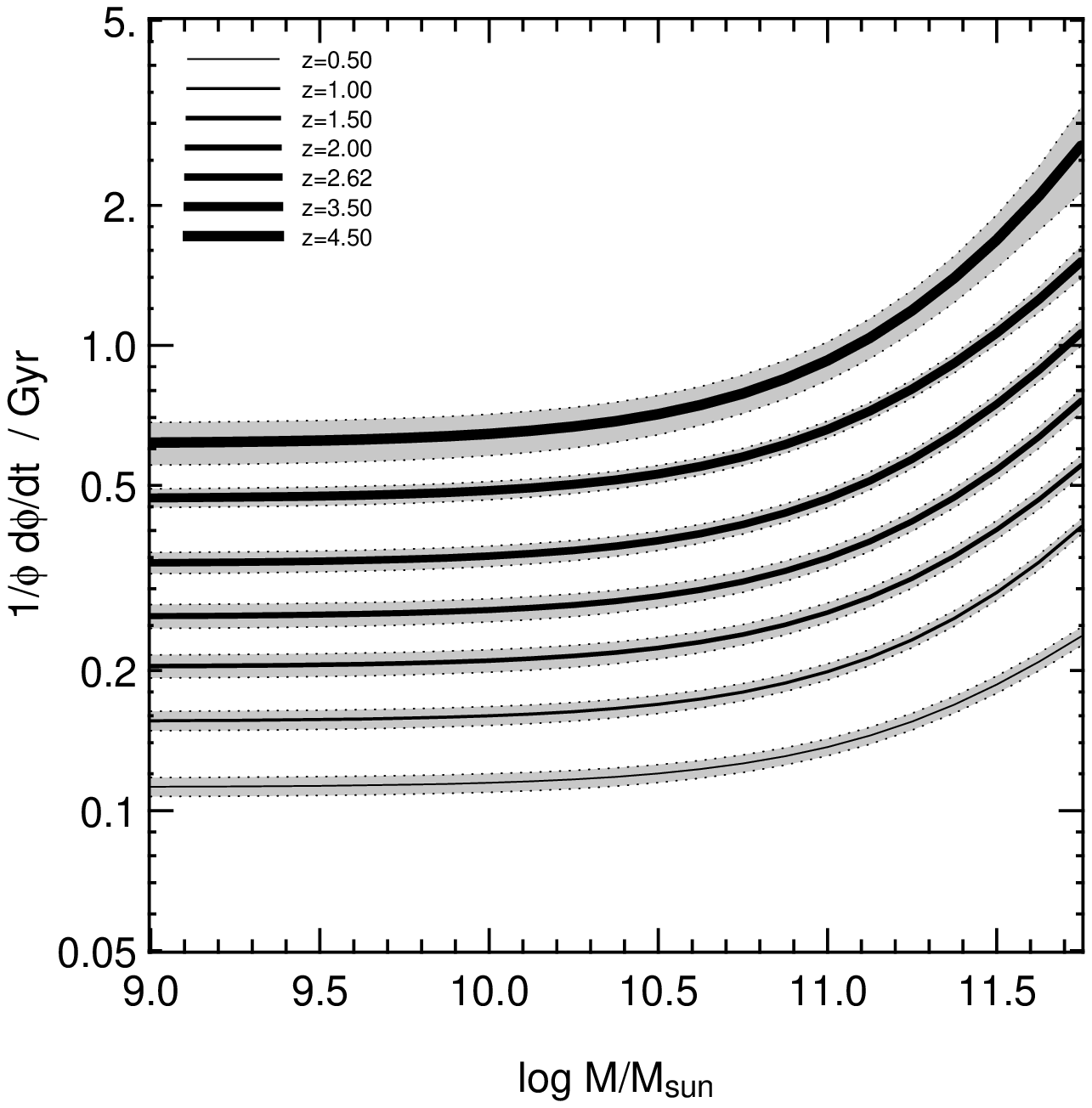}\hspace*{0.3cm}
  \includegraphics[width=0.45\textwidth]{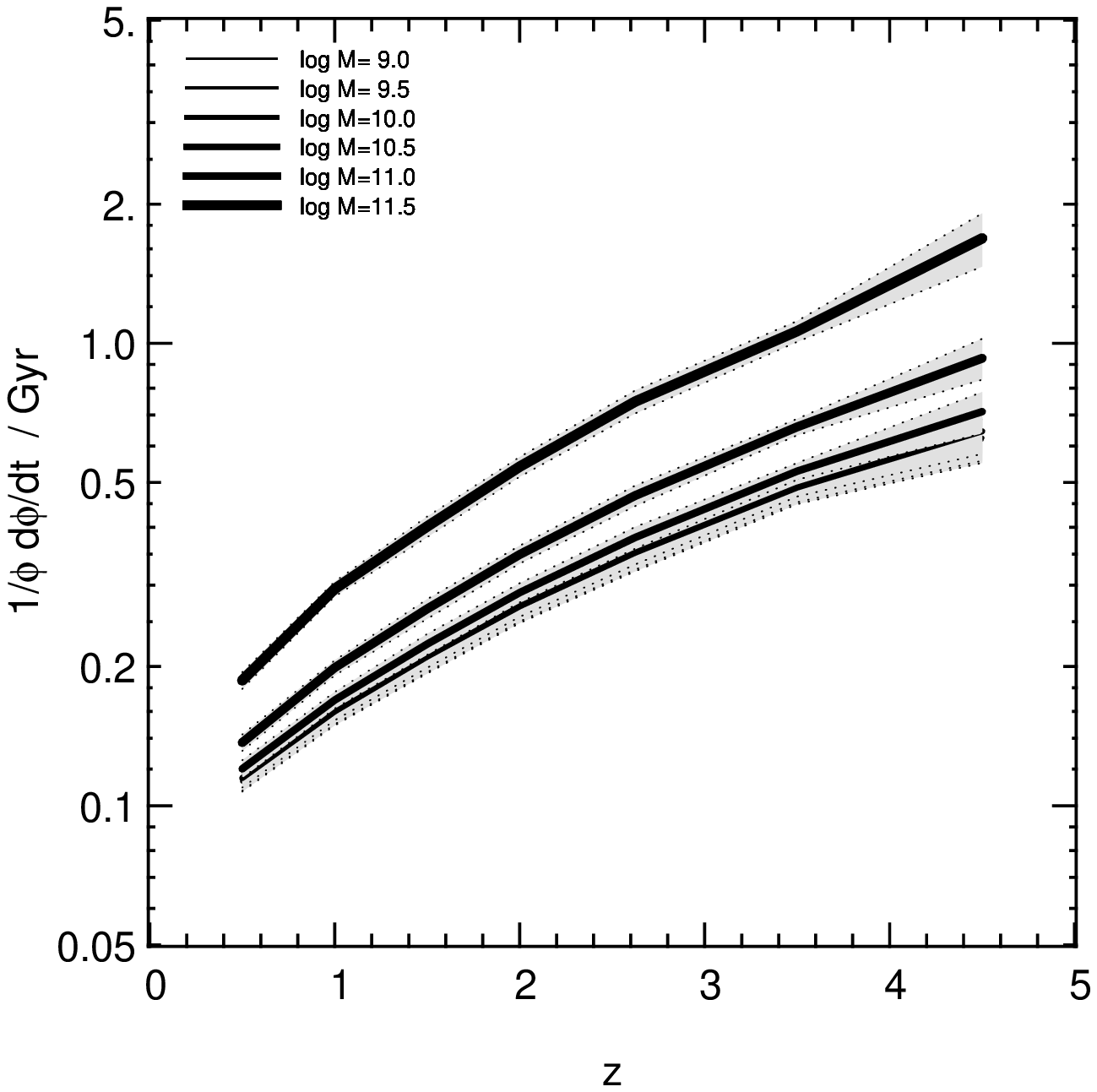}
  \caption{The partial derivative with respect to time of the stellar
    mass function as measured from the parametrization of its redshift
    evolution (see text). The left hand panel shows the derivative as
    a function of stellar mass at different redshifts; thicker lines
    indicate higher $z$ with values between $z=0.5$ and $z=4.5$ as in
    Fig.~\ref{fig:sfr}. The right hand panel shows the same functions
    but plotted against redshift, with higher stellar mass indicated
    by thicker lines; the mass values are indicated and the same as in
    Fig.~\ref{fig:sfr}. At masses below $M^*$, where the mass function
    is a power law, the evolution becomes independent of
    mass. Therefore curves for masses below $\log\Mstar \sim 10$ in
    the right panel lie on top of each other.\label{fig:dphidt}}
\end{figure*}

It has long been thought that star formation histories of galaxies are
roughly exponentials on average, with later type and lower mass
galaxies having longer $e$-folding times
\citep[e.g.][]{Sandage89}. From investigations of the present day
stellar population in elliptical galaxies, this was confirmed by
\citet{TMBM05}, at least for spheroidals. \citet{HPJD04} use spectra
of galaxies from the SDSS to infer star formation histories of
galaxies as a function of mass and find similar trends.  Here,
Fig.~\ref{fig:avgsfh} -- from measurements of star formation rates as
a function of stellar mass and redshift -- directly shows that a
relationship between star formation history and galaxy mass
holds. Rise time, peak star formation rate, peak time, and
post-maximum (exponential) decay timescale are all correlated with
mass.  More massive galaxies have steeper and earlier onsets of star
formation, higher peak star formation rate, and the following
exponential decay has a shorter $e$-folding time. Less massive
galaxies have a more gradual onset of star formation, a broader peak
of star formation activity, and their maximum star formation rate is
lower and occurs later. Their subsequent exponential decline has a
longer $e$-folding time.

Since the star formation rate as a function of mass is well represented
by a power law in mass with constant slope, we can write the early
star formation rate (at masses below the break mass) as $\dd M/\dd t =
A(t)\, M^{\beta}$, with $A(t) \propto \Mdot^0$ from
eq.~(\ref{eq:sf-parametrization}), and monotonically decreasing with
time. Hence, galaxies which begin forming stars later in time, have
longer rise times, reach lower peak star formation rate, and also
lower final mass. The latter statement can be derived from the fact
that the break mass decreases with time, and so galaxies that start
forming stars later have their star formation quenched at lower mass
than galaxies that start forming stars early, when the break mass is
still high.

As a side note, the existence of the curious universal power-law
relation between star formation rate and stellar mass can be explained
using disk galaxy scaling relations and the Schmidt-Kennicutt star
formation law \citep{KennicuttARAA98}. If one takes the
Schmidt-Kennicutt law to read $\sigsfr \propto \Sigma /
\tau_\mathrm{dyn}$ (with $\Sigma$ denoting surface density and $\tau$
being the dynamical timescale, see \citealp{Silk97,Kennicutt98}), then
$\Mdot \propto R^2\,\sigsfr \propto R^2\,\Sigma\,V/R \propto
\Sigma\,V\,R$ (with $V$ denoting the characteristic velocity of the
system and $R$ its radius). Using the virial relation we have $\Mdot
\propto \Mstar^{0.5}\,R^{0.5}$, which, with $M \propto R^3$, becomes
$\Mdot \propto \Mstar^{0.62}$. Using the disk scaling relations given
in \citet{MMW98} instead of the (halo) virial relation ($M \propto
V_c^3$; $R_d \propto V_c$), one obtains $\Mdot \propto \Mstar$ for
average halo spin parameter and fraction of disk angular
momentum. Both of these arguments assume a fixed scaling of cold gas
surface density with total mass surface density.


\section{The Time Derivative of the Mass Function}
\label{sec:dphidt}

Now that we have assembled all the necessary data, we turn back to
eq.~(\ref{eq:merge}).  The right hand side of eq.~(\ref{eq:merge}) has
two terms. The first is the observed time derivative of the mass
function, the second one is the change with time induced in the mass
function due to star formation. Here we will deal with the first term.

We use the parametrization of the evolution of the stellar mass
function as established in \S~\ref{sec:mass-function}. From this
parametrization, it is a straight forward procedure to calculate the
time derivative $\dpt \phi/\dpt t$. For convenience, we will
henceforth always measure time derivatives in terms of relative
changes, $(1/\phi)\,(\dpt \phi/\dpt t)$. These are plotted against
mass in the left hand panel of Fig.~\ref{fig:dphidt}; we use thicker
lines to indicate higher redshift. The right hand panel shows the same
quantity, this time plotted against redshift, with line thickness
indicating increasing stellar mass. We plot data for the same
redshifts and masses as in Fig.~\ref{fig:sfr}.

The shaded regions indicate the 1-$\sigma$ confidence limits
calculated using Monte Carlo simulations of the fit of the evolution
of the mass function, eqs.~(5) -- (7) and Fig.~\ref{fig:schechter} in
\S~\ref{sec:mass-function}. It is important to note that while the
uncertainties on the mass function itself are rather large, the
parameterization of its evolution provides much smaller formal
uncertainties on the derivative with respect to time that is of
interest here.

\begin{figure*}[t]
  \centering
  \includegraphics[width=0.45\textwidth]{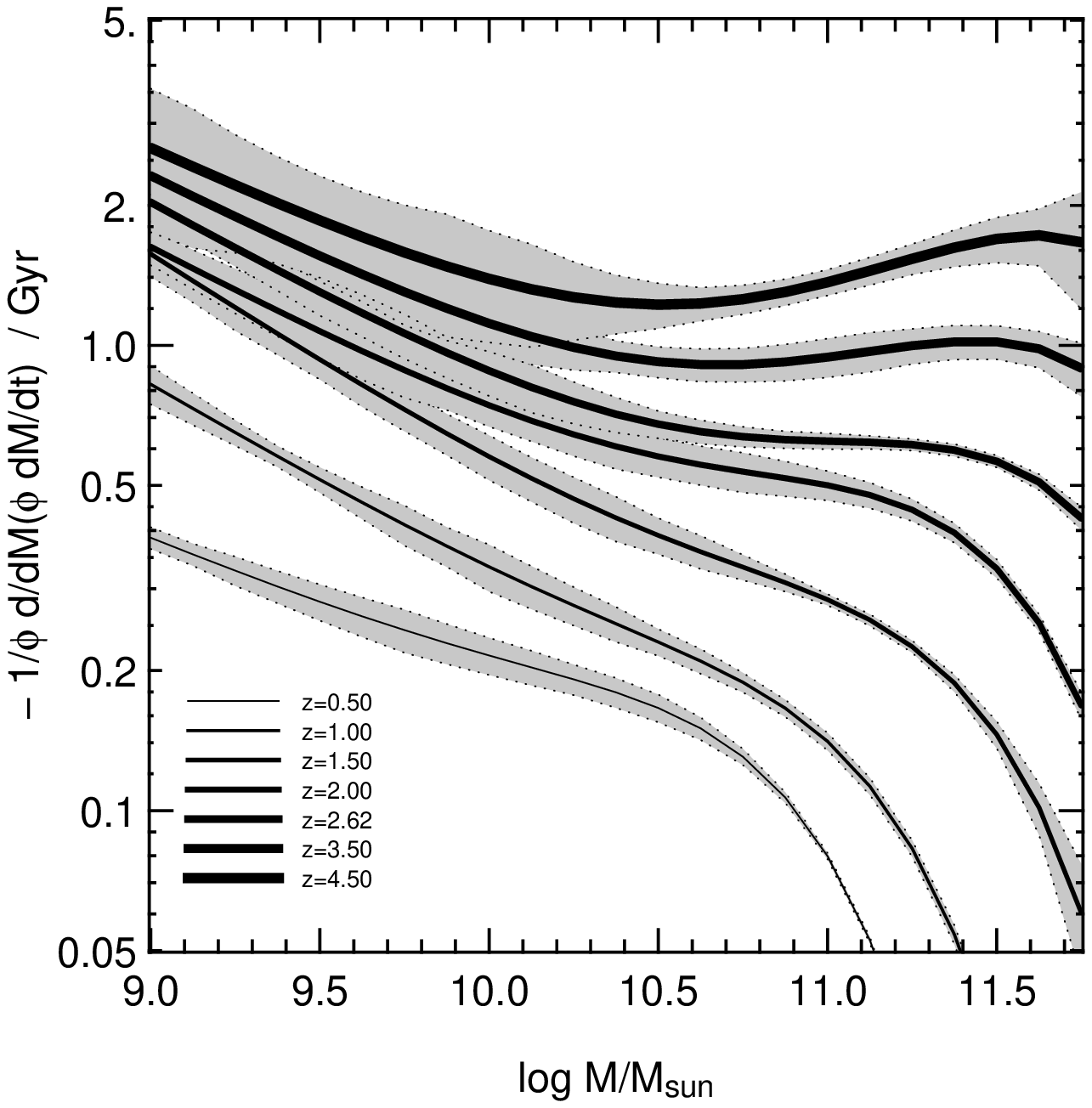}\hspace*{0.3cm}
  \includegraphics[width=0.45\textwidth]{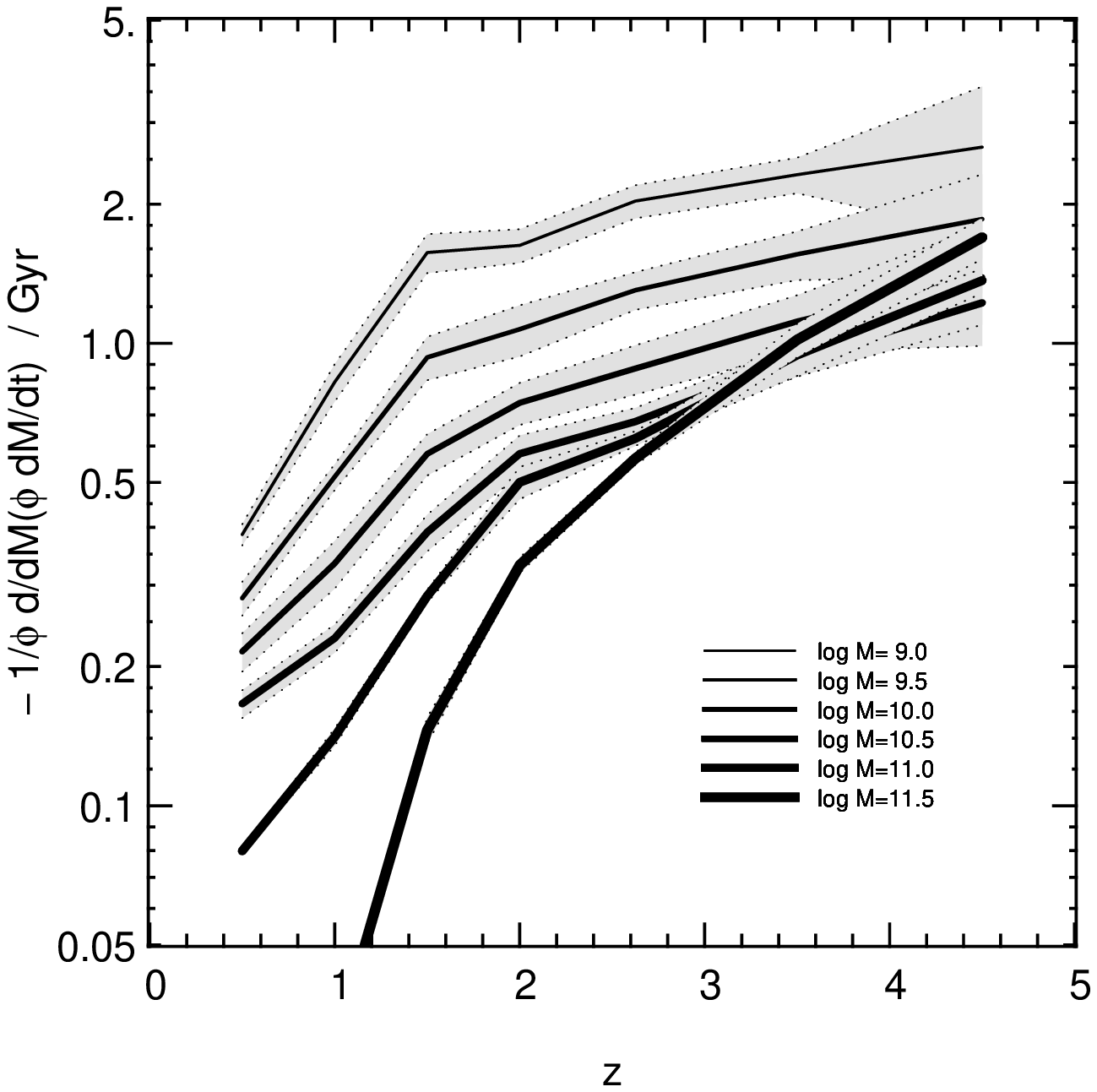}
  \caption{The relative change per Gyr in the galaxy stellar mass
    function due to star formation only, plotted against mass (left
    panel) and redshift (right panel). These curves are the result of
    evaluating eq.~\ref{eq:cont}. Confidence limits are derived by
    directly propagating the confidence limits on the average star
    formation rate through eq.~\ref{eq:cont}. \label{fig:dphisfr}}
\end{figure*}

Note that our parametrization of the mass function assumes a fixed
faint-end slope of $\alpha = -1.3$ (the data do not allow us to
measure the slope at higher redshift; up to $z \sim 2$, the slope is
consistent with being constant). Hence the time derivative of the mass
function becomes independent of mass at masses below the
characteristic mass, $M^*(z)$, where the mass function is a power
law. In the right hand panel of Fig.~\ref{fig:dphidt} this means that
the low mass curves (below $\log\Mstar \sim 10$) all lie on top of
each other.

The most rapid buildup of the mass function happens (not unexpectedly)
at high redshift and at the high-mass end. At $z \sim 4$ and
$\log\Mstar \sim 11$, the number density approximately doubles every
Gyr, while the density of $\log\Mstar \sim 10$ objects (typical Lyman
break galaxies) increases by 50\% per Gyr. At $z \lesssim 1$, the
change in number densities is much more modest, of the order of
10-20\% per Gyr (and likely decreasing further rapidly), so that the
mass density roughly doubles between $z \sim 1$ and $z = 0$, a
behavior observed and confirmed by many studies mentioned in the
introduction.

There is rather little differential evolution in the Fors Deep Field
stellar mass function in the sense that the ratio of the change at
high mass and at low mass evolves from a factor of $\sim 4$ at high
$z$ to a factor $\sim 2$ at $z \sim 0.5$. This has been noted in
\cite{FDFGOODS05}; see their Fig.~4. \citet{Fontanaetal06}, analyzing
data similarly in the GOODS-MUSIC survey, find that this ratio tends
to zero at late times, such that the buildup of the $\log\Mstar > 11$
galaxy population is complete by $z \sim 1$ while it still continues
slowly in our sample (compare also studies of elliptical galaxies,
e.g.~\citealp{VdWetal05}). This discrepancy, apparent in a number of
datasets -- whether it is due to the small volume of the FDF at low
$z$, the photometric redshifts, the stellar population modeling
technique, or other effects -- has yet to be resolved.

Since the main focus of this paper is to discuss the method presented
here to analyze mass function and star formation rate data, we will
not spend too much time discussing peculiarities of the present
dataset. We are fully aware that the current data have significant
deficiencies (both in mass and star formation rate), yet they still
offer the possibility to discover and learn about fundamental trends
in the formation and evolution of galaxies.


\section{The Evolution of the Mass Function due to Star Formation}
\label{sec:dphisfr}

After the derivative of the mass function, we turn to the second term
on the right hand side of eq.~\ref{eq:merge}. We proceed to analyze
the relative change in the stellar mass function that is due to star
formation only, $\psi \equiv \frac{1}{\phi}\,\frac{\dpt \pmt}{\dpt t}
\Big\vert_{\mathrm{SF}}$, which is $\phi^{-1}$ times
eq.~\ref{eq:cont}. We will use the symbol $\psi$ to refer to this
quantity in what follows.

To compute the function $\psi = \phi^{-1}\,(-\dpt/\dpt\,\Mstar)\,\,
(\phi \, \Mdot)$, in each time (redshift) bin, we use the
parameterized fits to $\phi(\Mstar,t)$ and $\Mdot(\Mstar,t)$ listed in
Tables~\ref{tab:mffit} and \ref{tab:sfrfit}, respectively. We plot the
resulting curves in Fig.~\ref{fig:dphisfr}. The left hand panel shows
the relative change of the stellar mass function plotted against
stellar mass in seven redshift bins spanning $0.5 < z < 4.5$; as in
the previous plots, thicker lines indicate higher redshift. The right
hand panel shows the same data but now plotted against redshift with
stellar mass coded as line thickness. Masses and redshifts are the
same as in Fig.~\ref{fig:dphidt}. Confidence regions are derived
directly from the confidence limits on the average star formation
rates propagated through eq.~\ref{eq:cont}.

$\psi(\Mstar,t)$ is positive at all times. This means that the number
density of galaxies is predicted to increase with time due to the star
formation activity of the galaxies. Note that this is not a trivial
statement: If the number of objects leaving a mass bin $[\Mstar,
\Mstar+\dd\Mstar)$ in the time interval $\dd t$ due to their star
formation rate $\Mdot(\Mstar,t)$ is larger than the number of objects
entering the same bin from lower mass due to their own star formation
(see eq.~(\ref{eq:cont})), the number density at this mass would
decrease. It is a matter of comparing the derivatives w.r.t.\ mass
of the mass function and the star formation rate.

In the low-mass range, the stellar mass function is well represented
by a power law with slope $\alpha \sim -1.3$ (locally), and we have
assumed that this slope does not evolve with time (which is backed by
the data up to $z \sim 2$; see \S~\ref{sec:mass-function} and also
\citealp{Fontanaetal06}). The star formation rate is also well
described by a power law (see \S~\ref{sec:sfr}), $\Mdot \propto
\Mstar^\beta$, with $\beta \sim 0.6$. Eq.~(\ref{eq:cont}) times
$\phi^{-1}$ hence reduces to
\begin{equation}
  \psi \propto - \frac{1}{\Mstar^{\alpha}}\, \frac{\dpt}{\dpt \Mstar}\,
  \left[ \Mstar^{\alpha}\, \Mstar^\beta \right]
  \; \propto \; -(\alpha+\beta)\,\Mstar^{\beta-1}\, .
\end{equation}
This term is positive if $\alpha + \beta < 0$.  With $\alpha \sim
-1.3$ and $\beta \sim 0.6$, this condition is satisfied at all times
at the faint end. At the bright end, the mass function steepens
quicker than the star formation rate does, and the condition remains
satisfied.

We wish to note here that we are very likely to miss low-mass low star
formation part of the galaxy population, hence we very likely
underestimate the slope of the stellar mass function, especially at
higher redshift.

At high redshift and high mass, we see that star formation leads to a
strong increase in number density. The number of objects with $\log
\Mstar \gtrsim 11$ roughly doubles every Gyr at $z\sim 4$. This rate
of buildup quickly decreases with time, slowing down to an increase by
50\% per Gyr by $z \sim 2.5$. At $z \lesssim 1.5$, star formation has
dropped to contribute a relative increase in number density of $\sim
0.1$~Gyr$^{-1}$, and to 0.05~Gyr$^{-1}$ at $z \sim 0.5$.

At low masses, $\log\Mstar \lesssim 10$, the number density very
rapidly increases (due to star formation) at a rate comparable to that
of high-mass objects. Unlike the high mass objects, however, the
growth due to star formation persists in low mass objects for much
longer, dropping to a rate of 0.5~Gyr$^{-1}$ at $z \sim 1$ and
0.2~Gyr$^{-1}$ at $z \sim 0.5$. While the number density of massive
galaxies is not changing significantly anymore due to star formation
at low $z$, that of low-mass galaxies does.


\section{The Assembly Rate}
\label{sec:mergerrate}

\begin{figure*}[t]
  \centering
  \includegraphics[width=0.45\textwidth]{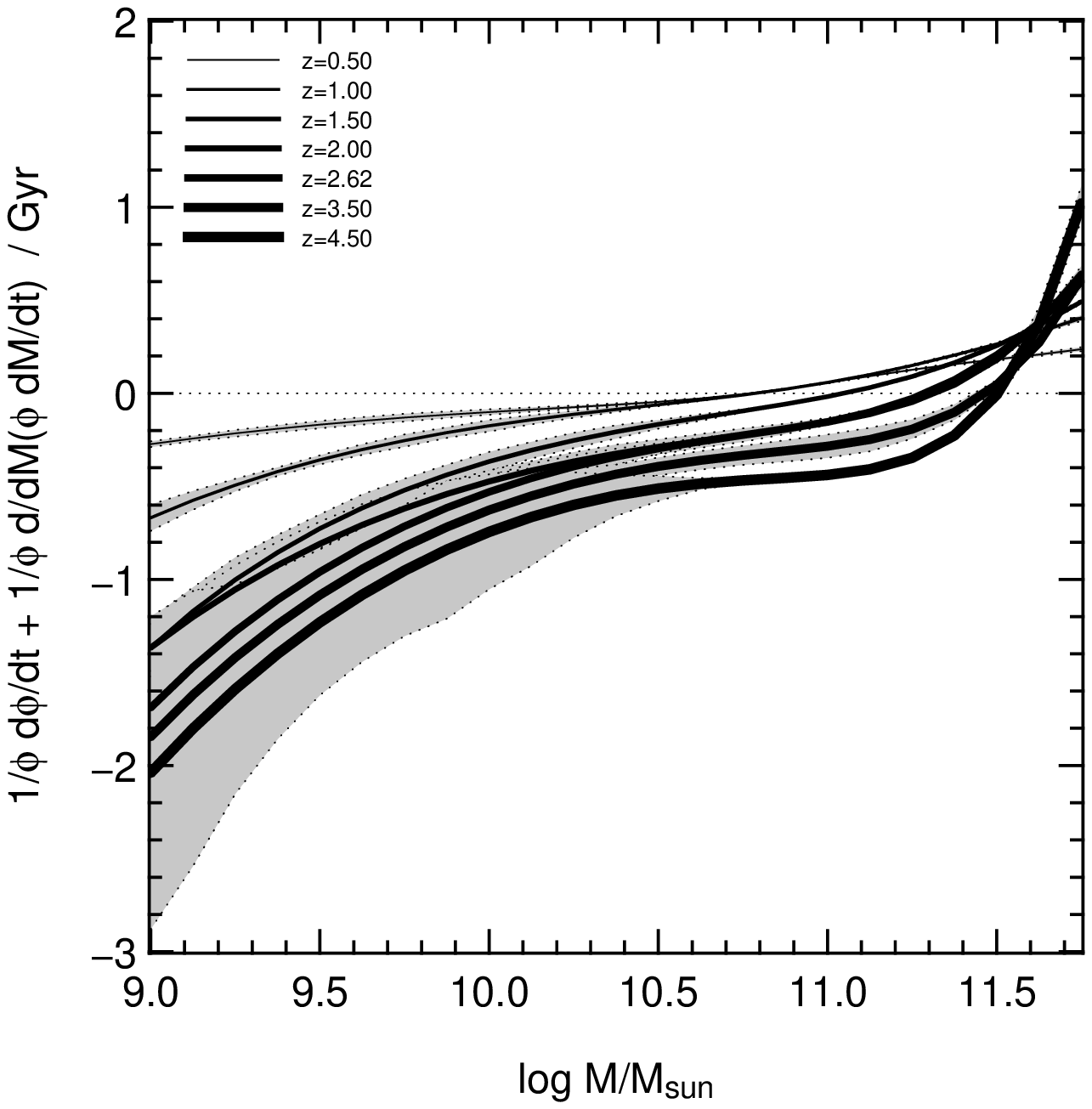}\hspace*{0.3cm}
  \includegraphics[width=0.45\textwidth]{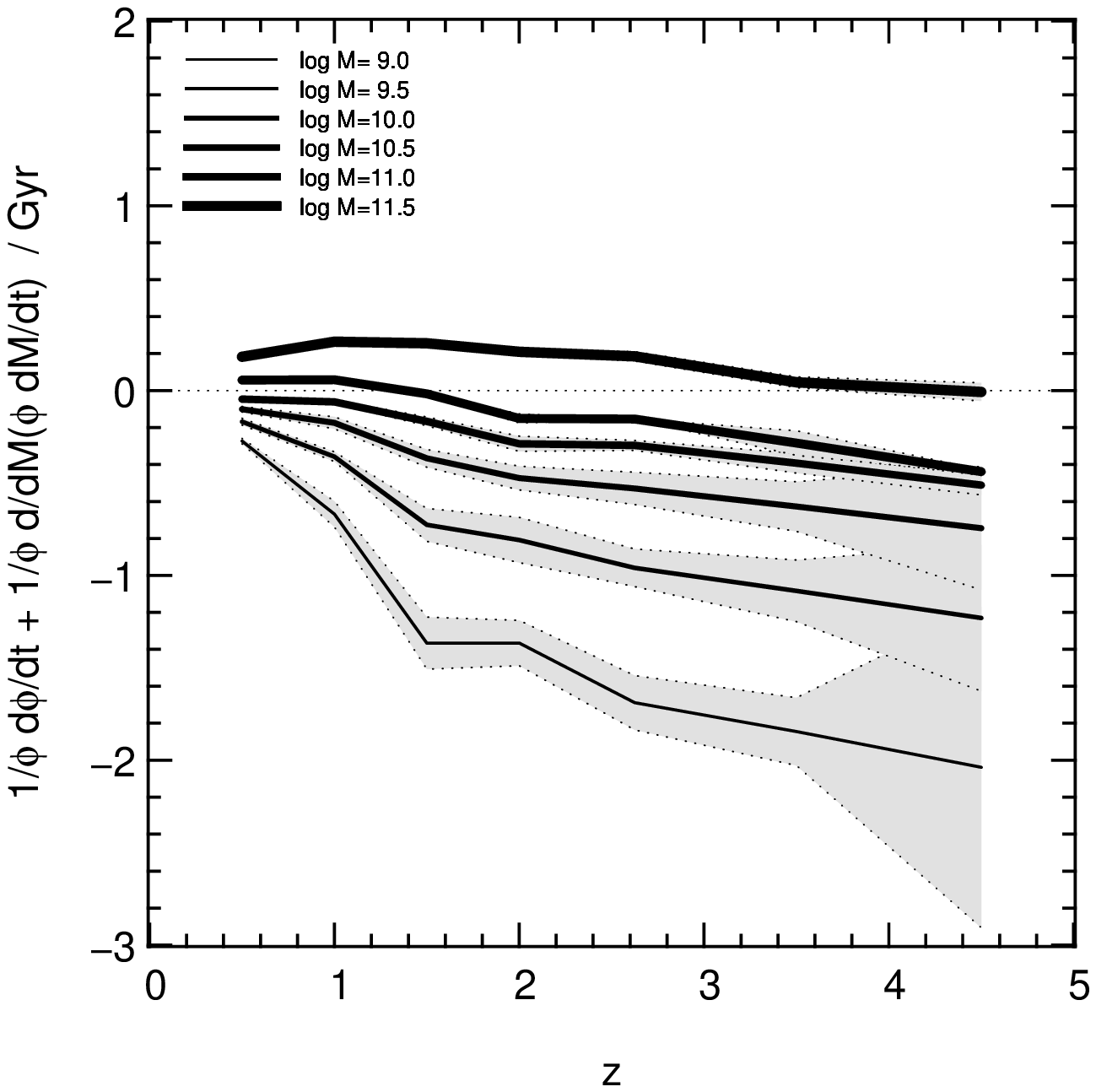}
  \caption{The {\em assembly rate}: the relative change in the galaxy
    stellar mass function that is caused by anything other than star
    formation. We interpret this quantity as being due to accretion
    and merging of galaxies. The left hand panel is plotted against
    mass, the right hand panel against
    redshift.\label{fig:mergerrate}}
\end{figure*}

Finally, we have all the ingredients necessary to fully evaluate
eq.~(\ref{eq:merge2}), which we do in Fig.~\ref{fig:mergerrate},
showing our results again versus mass in the left panel and against
redshift in the right panel.  We will use the less cumbersome term
{\em assembly rate} for ``change in number density due to merging'' in
what follows. We remind the reader to be mindful of the difference
between this quantity and the merger rate (see
\S~\ref{sec:desc-mass}).

Confidence regions in Fig.~\ref{fig:mergerrate} are calculated by
means of Monte Carlo realizations of data within the uncertainties in
the mass function evolution and the star formation rates, and
propagating those through eq.~(\ref{eq:merge2}). It is worth noting
that since the constraints on the derivatives of $\phi^*$ and $\log
M^*$ with respect to time are much tighter than the constraints on
these quantities themselves, the error budget in
Fig.~\ref{fig:mergerrate} is dominated by the uncertainty in the
average star formation rate. This leads to formally small errors at
masses and redshifts with low star formation rates.

The curves in Fig.~\ref{fig:mergerrate} are negative at low mass and
positive at high mass, with the zero-crossing mass evolving from
higher masses at high redshift to lower masses at lower redshift. This
means that once star formation is accounted for, a source of high mass
objects and a sink of low mass objects is necessary to explain the
mass function data. In the most general terms, this is consistent with
merging activity building up higher mass objects from (many) lower
mass ones. In this picture, the merger rate decreases with time (the
absolute value of the curves decreases with time). Also, the mass at
which the merger rate tends to increase the number density instead of
decreasing it (the mass above which objects are accreted onto and
below which objects are destroyed) decreases with time. However, an
important caveat is that a loss of objects could also be due to
(faint) objects dropping out of the survey as soon as their star
formation activity ceases. This can partly explain the negative values
observed at low masses and the trend mentioned above.

In the following we shall look into the merger rates as a function of
mass and time in more detail.

\subsection{High redshift and high mass}

At high redshift and high masses, the time-evolution in the mass
function shows a rapid increase in number density of high mass objects
(see \S~\ref{sec:dphidt}).  Star formation accounts for most of this
increase, leaving a difference consistent with zero at $\log\Mstar
\sim 11$ and a small positive difference of $0.2-0.8$~Gyr$^{-1}$ at
$\log\Mstar \sim 11.5$ and $z \gtrsim 3$.  However, our
parametrization of $\phi^*(z)$ flattens at high redshifts and
over-predicts the normalization of the mass function in the bin at high
$z$ (compare the high-$z$ data point in the upper panel of
Fig.~\ref{fig:schechter}).  As a result, the total change in the mass
function is smaller in the parameterized fit to $\phi^*(z)$ compared
to the raw data by about 30\%. Therefore, if we use the Schechter
functions shown in Fig.~\ref{fig:mf} directly, which show a steeper
increase in $\phi^*$ at $z \gtrsim 3$, instead of the smooth
parametrization $\phi^* \propto (1+z)^{-1.07}$, the curves in
Fig.~\ref{fig:mergerrate} move up, and the inferred assembly rate
becomes larger by the same amount.

Given this uncertainty, we feel that the star formation rate we
observe is roughly sufficient to explain the increase in number
density at high mass and high redshift. As an upper limit, at most
0.5--0.8 effective major mergers per Gyr are consistent with the
data. This number, though, would be enough to transform most of these
early high mass objects into ellipticals contemporaneously with their
major star formation episode. This is consistent with the picture
emerging from various observations in the local universe and at high
redshift which show that most massive ellipticals form quickly and
early, with more massive galaxies being older, and their stars having
formed more rapidly (e.g.\
\citealp{BZB96,Treuetal01c,VDF03,TMBM05,VdWetal05,Renzini06}).

\subsection{High redshift and low mass}

As we have seen, the data point towards small galaxies being destroyed
and large galaxies being built by mergers, with the caveat of
selection effects being responsible for parts of this trend.

A plausible interpretation instead of invoking a high destruction rate
due to mergers, is based on the fact that low mass objects with low
star formation rates (and therefore redder colors) are very likely
missing from the survey. If the slope of the mass function were
steeper due to the existence of a class of low-mass objects missing
from the survey, we'd need an even higher destruction rate at low
masses, due to the increase in numbers and decrease in the average
star formation rate. However, if the star formation rate as a function
of mass were steeper, $\Mdot \propto \Mstar^{0.8}$ instead of $\Mdot
\propto \Mstar^{0.6}$ either due to lower star formation at low masses
or higher star formation at high masses, the difference would be close
to zero below $\log\Mstar \sim 11$ at all times, and little change
($\sim 0.2$~Gyr$^{-1}$) in number density due to merging would be
necessary. This would imply that at low masses, merging is creating
objects of any particular mass at the same rate as these objects are
incorporated into larger structures.

\subsection{Low redshift and high mass}

Our current dataset is most reliable (both in terms of mass
completeness and robustness of the star formation rate estimate) at
redshifts $z \lesssim 2$ and at masses $\log \Mstar \gtrsim
10.5$. Here, the photometric data are deep enough to detect objects
over a large dynamic range in $\Mstar/L$, and, also, intergalactic
absorption (the Lyman break) does not affect the optical bands yet,
such that the object's spectral energy distributions are well-sampled
from the rest-frame UV into the optical and near-IR.

At $z \lesssim 2.5$ the assembly rate is increasing towards higher
stellar masses and towards lower redshift, mostly due to the rapid
decline of the star formation rate in this regime, but it is always
very low. Most of the increase in number density of massive galaxies
is fully explained by the star formation rates observed.

It is very difficult to measure the high mass end of the mass function
accurately. These objects are rare, and on the steep exponential tail
of the mass function any small error in distance (due to the use of
photometric redshifts) or error in $\Mstar/L$ will lead to a large
error in the number density. Also, the effects of cosmic variance are
largest in this mass range and may heavily bias results especially in
relatively small-area surveys as the FDF. Some assurance is provided
by the fact that the $z\sim 1$ mass functions of a number of surveys
agree reasonably well.  If our measurement of the high mass end of the
mass function at $z < 1.5$ is correct, mergers only contribute to the
growth of the mass function at $\log \Mstar \gtrsim 11$ and only on a
low level. A growth of only 0.1 -- 0.2~Gyr$^{-1}$ is consistent with
our data. This is equivalent to about 1 effective major merger per
object in the redshift range $1.5 > z > 0$. If, on the other hand,
there is less evolution at the high mass end of the mass function,
e.g.\ as found by \citet{Fontanaetal06}, then no merging at all would
be required at high mass (compare also the mass functions in
\citealp{K20-04} and \citealp{MUNICS6}). However, there is strong
observational evidence that recent merging does play a role at least
in field elliptical galaxies (e.g.\
\citealp{Schweizer82,WMSF97,GMKMM01,Genzeletal01}).

Other studies of the merger rate do not help to answer this
question. On one side, \citet{vD05} studied signatures of interaction
and tidal debris in a sample of nearby red galaxies. With some
assumptions on the timescale of the merger events, they conclude that
the present day accretion rate in E/S0 galaxies is $\Delta M/M =
0.09\pm 0.04$~Gyr$^{-1}$, a value that is fully consistent with our
measurement, assuming that red sequence galaxies comprise the majority
of the $\log \Mstar > 11$ local galaxy population. Also,
\citet{Belletal06b} find that 50-70\% of galaxies with $\log \Mstar >
10.7$ have undergone a major merger since $z \sim 1$ (see also
\citealp{Belletal06a}). On the other side, \cite{Linetal04} and
\citet{Lotzetal06} find a much lower merger rate. They argue that in
the DEEP2 survey, only $\sim 10\%$ of massive galaxies have undergone
a major merger since $z \sim 1.2$. \citet{Bundyetal06} also argue for
a low contribution of mergers at the high mass end. An overview of a
number of studies is given in \citet{Kartaltepeetal07}.

\subsection{Evolution with time}

As we have mentioned, the general trend in the data is for the
assembly rate to become more positive with time. The zero-crossing
mass decreases with time. In high mass objects, the assembly rate
starts out at zero or slightly below, with star formation rates
dominating the initial increase in number density. The assembly rate
then increases, such that at low $z$, merging becomes the dominant
process that increases the number density at masses $\log\Mstar
\gtrsim 11$. The growth rate in this mass and redshift range is 0.1 --
0.2~Gyr$^{-1}$. Lower-mass galaxies are consumed in these mergers. The
mass at which the assembly rate transitions from objects being consumed
to objects being formed decreases with time, roughly tracing the break
mass in the star formation rate. The zero-crossing of the assembly rate
occurs at $\log \Mstar = 10.8$ at $z = 0.5$, $\log \Mstar = 11.0$ at
$z = 1.5$, $\log \Mstar = 11.3$ at $z = 2$, and $\log \Mstar = 11.5$
at $z = 3.5$. While merging is still happening below this transition
mass, the net number of objects decreases there. Hence, at lower
redshift, the number density of progressively lower-mass galaxies
starts to increase due to merging. This is a direct observation of the
top-down buildup of the red sequence, where major merger remnants are
found \citep[e.g.][]{Belletal2004,Faberetal05,Bundyetal06,Belletal07}.

It is interesting to note that in CDM, e.g.\ in the \citet{PS74}
formalism, we expect the opposite behavior: the characteristic mass
increases with time, and hence progressively more massive halos
preferably get incorporated into even larger structures, and their
number density begins to decline. However, this formalism assumes that
merged halos do not survive as substructure within their parent
halos. Accounting for the survival of substructure, e.g.\ the
existence of satellite galaxies or galaxies within clusters, might
account for this apparent discrepancy. This will be explored in a
follow-up paper.


\section{Summary, Discussion, and Conclusions}
\label{sec:summary}

Summarizing our results, we present average star formation rates as a
function of stellar mass and redshift, $\Mdotmt$, for galaxies in the
Fors Deep Field spanning $0 < z < 5$. We directly measure their star
formation histories as a function of stellar mass by integrating the
former quantity.

At $z \gtrsim 3$ the average star formation rate is monotonically
rising with stellar mass and is well-fitted by a power law of stellar
mass with exponent $\beta \sim 0.6$. The average star formation rates
in the most massive objects at this redshift are 100 -- 500~$\Msun
\mathrm{yr}^{-1}$, suggesting that these are the progenitors of modern
day massive ellipticals seen during their major epoch of star
formation. At redshift $z \sim 3$, the star formation rates starts to
drop at the high mass end while still being an increasing function of
mass at lower masses. The power-law exponent is consistent with
staying constant. As redshift decreases further, the star formation
rates drop at progressively lower masses (downsizing), dropping most
rapidly with redshift for high mass ($\log\Mstar \gtrsim 11$)
galaxies. The mass at which the star formation rate starts to deviate
from the power law form (break mass) progresses smoothly from $\log
\Mstar^1 \gtrsim 13$ at $z\sim 5$ to $\log \Mstar^1 \sim 10.9$ at
$z\sim 0.5$. For comparison, in the local universe
\citet{Kauffmannetal03b} find a characteristic mass of $\log \Mstar
\sim 10.4$ at which the galaxy population changes from actively star
forming at lower stellar mass to typically quiescent at higher masses.
The break mass is evolving with redshift according to $\Mstar^1(z) =
2.7\times 10^{10} (1+z)^{2.1}$.

We directly show that a relationship between star formation history
and galaxy mass holds. Rise time, peak star formation rate, peak time,
and post-maximum (exponential) decay timescale are all correlated with
mass.  More massive galaxies have steeper early onsets of star
formation, their peak star formation occurs earlier and is higher, and
the following exponential decay has a shorter $e$-folding time.

We then present a method to infer effective merger rates (or mass
accretion rates) from observations of the galaxy stellar mass function
and the star formation rate as a function of mass and time. We
quantify the evolution of the mass function due to the star formation
activity of galaxies. We compare this evolution to the observed time
derivative of the mass function and interpret any difference between
the two as due to changes in galaxy mass due to external causes;
accretion and merging are dominant among such processes. We introduce
and analyze a quantity that we call the {\em assembly rate}, which is
defined as the change in number density of galaxies due to accretion
and merging.

We observe sufficient star formation to explain the increase in number
density at high mass and high redshift. As an upper limit, at most 0.8
effective major mergers per Gyr are consistent with the data. This
number, though, would be enough to transform most of these high mass
objects into ellipticals contemporaneously to their major star
formation episode.

The data point towards small galaxies being destroyed and large
galaxies being built by mergers, with the caveat of selection effects
being responsible for parts of this trend.  A plausible
interpretation, instead of invoking a high destruction rate due to
mergers, is based on the fact that low mass objects with low star
formation rates (and therefore redder colors) are very likely missing
from the survey. The general trend in the data is for the assembly
rate to become more positive with time. The zero-crossing mass
decreases with time. In high mass objects, the assembly rate starts
out at zero or slightly below, with star formation rates dominating
the initial increase in number density. The assembly rate then
increases, such that at low $z$, merging becomes the dominant process
that increases the number density at masses $\log\Mstar \gtrsim 11$. A
growth of 0.1 -- 0.2~Gyr$^{-1}$ is consistent with our data at $z
\lesssim 1.5$. This is equivalent to about 1 effective major merger
per object in the redshift range $1.5 > z > 0$.

Synthesizing the above into a picture of the formation of massive
ellipticals, a picture can be drawn in which elliptical galaxies would
have formed their stars quickly and early ($z \gtrsim 2.5$) and
subsequently quenched their star formation, with the highest mass
galaxies forming first, and on the shortest timescale. Their
elliptical morphology may be a consequence of early merging activity
given the merger rate that is consistent with our data at high
redshift ($\sim 0.8$~Gyr$^{-1}$; see above). Later, these galaxies may
continue to evolve along the red sequence, however during this phase
at $z \lesssim 1.5$, not much growth by merging is happening any more
(at most a factor of two in stellar mass).

At more moderate masses of $10 < \log\Mstar < 11$, we find that the
evolution of the number density due to mergers points towards these
galaxies being preferably destroyed at early times, while at later
times the change in their numbers turns positive. At lower redshift,
the number density of progressively lower mass galaxies starts to
increase due to merging. This is a direct confirmation of the top-down
buildup of the red sequence suggested by recent observations (e.g.\
\citealp{Belletal2004,Faberetal05,Bundyetal06,Belletal07}) and the
simultaneous shutdown of star formation in these objects.

The results presented here do depend on some assumptions that we had
to make, mainly due to general deficiencies of current observations
and more specifically of the dataset we have at hand.  Assuming that
the faint-end slope of the MF does not evolve is consistent with the
observations only up to $z \lesssim 2$
\citep{FDFGOODS05,Fontanaetal06}. A steeper faint-end slope is
possible due to faint objects (with lower star formation rates than
the detected ones) being missed due to the survey flux
limit. Therefore, a steeper faint-end slope of the MF also means a
flatter average star formation rate, and hence a larger assembly rate
at low masses.

The fitting function one chooses to use to describe the evolution of
the mass function also affects the results. For example, a power-law
fit to the evolution of $\phi^*$ constrains the derivative of the MF
w.r.t.\ time, and thus limits the possible range of assembly
rates. Ideally, one would like to have MF data robust enough to be
used to measured derivatives directly and avoid the use of fitting
formulae entirely.

The choice of a single average value for the SFR at each mass and
redshift also affects the results. In reality, galaxies show a range
of activity at fixed mass. Allowing for such a range of possible
values instead of a single number would lend the method more
flexibility, but also demands much higher quality star formation data
with better-understood selection effects.

The current generation of ongoing large-area surveys offer the
possibility to further explore the build up of the red sequence. With
a much larger number of red and massive objects, one could improve on
the analysis presented above in various ways. Studying merger
processes as a function of the environment is one such
possibility. Another promising approach is to separate galaxies into
passive objects and objects that are actively forming stars. One can
then apply our method to these two classes of objects individually,
adding another term to eq.~(\ref{eq:merge2}) that describes the rate
of migration between the two sequences. This will allow to investigate
the relation between merging activity and quenching of star formation
in a very direct way.


\acknowledgments

We wish to thank Peter Schuecker, Roberto Saglia, and Ralf Bender for
stimulating and encouraging discussions. We thank Ulrich Hopp for
commenting on an early version of the manuscript. We thank the
anonymous referee for providing valuable feedback.



\end{document}